\documentclass[runningheads,a4paper]{llncs}

\usepackage{amssymb}
\usepackage{graphicx}
\usepackage{url}
\urldef{\mailsa}\path|wolfgang.schmidt@gemy.de|
\urldef{\mailsb}\path|michael.hanspach@studium.fernuni-hagen.de|
\urldef{\mailsc}\path|joerg.keller@fernuni-hagen.de|
\newcommand{\keywords}[1]{\par\addvspace\baselineskip
\noindent\keywordname\enspace\ignorespaces#1}

\begin{document}

\mainmatter  

\title{A Case Study on Covert Channel Establishment via Software Caches in High-Assurance Computing Systems}

\titlerunning{A Case Study on Covert Channel Establishment}

%
%
\author{Wolfgang Schmidt, Michael Hanspach, J{\"o}rg Keller}%
%

\institute{FernUniversit{\"a}t in Hagen\\
\mailsa\\
\mailsb\\
\mailsc
}

%
%

\toctitle{Lecture Notes in Computer Science}
\tocauthor{Authors' Instructions}
\maketitle

\begin{abstract}
Covert channels can be utilized to secretly deliver information from high privileged processes to low privileged processes in the context of a high-assurance computing system.
In this case study, we investigate the possibility of covert channel establishment via software caches in the context of a framework for component-based operating systems.
While component-based operating systems offer security through the encapsulation of system service processes, complete isolation of these processes is not reasonably feasible.
This limitation is practically demonstrated with our concept of a specific covert timing channel based on file system caching.
The stability of the covert channel is evaluated and a methodology to disrupt the covert channel transmission is presented.
While these kinds of attacks are not limited to high-assurance computing systems, our study practically demonstrates that even security-focused computing systems with a minimal trusted computing base are vulnerable for such kinds of attacks and careful design decisions are necessary for secure operating system architectures.

\end{abstract}

\section{Introduction}\label{sec1}

The concept of a \emph{covert channel} has been described by Lampson as early as 1973~\cite{lampson:1973:ncp:362375.362389}.
Covert channels resemble potential means for communication that have not been designed for communication at all.
They can be applied by attackers to secretly extract data from computing systems.
In a different usage context, covert channels can also be applied to circumvent Internet censorship and other kinds of content restrictions~\cite{Feamster:2002:ICW:647253.720281}.

In this article, the implementation of covert channels for \emph{high-assurance} computing scenarios is discussed, in which the highest standards for security precautions are obligatory.
High-assurance computing systems are frequently used in the context of critical infrastructures and within high-security environments (e.g., in the military and intelligence context).
In difference to standard ``low-assurance'' computing systems, for high-assurance computing systems, ``\emph{convincing evidence is required that the system satisfies a collection of critical properties}''~\cite{Heimdahl98formalmethods}.

This evidence can be supplied via technical evaluation or formal verification, but this is usually much easier with component-based operating systems~\cite{jaeger:1998:sac:319195.319229}, which rely on a minimized TCB (trusted computing base).
In component-based operating systems, all system services are provided in isolated compartments, and a micro kernel serves as the central hub for IPC (inter-process communication).
As component-based operating systems heavily rely on the security of the operating system architecture, the design of the operating system architecture is specifically targeted by this article.

The contribution of this article includes considerations on the establishment of covert timing channels in the context of secure operating system architectures.
The covert channel presented in this article is a covert timing channel, which is based on different timing behaviors depending on the eviction status of a file system cache.
Furthermore, our work includes considerations on the mitigation of these kinds of covert channels, both in terms of disruption of covert channel transmissions, and in terms of preliminary prevention due to careful design of the operating system architecture.
This article is based upon the master's thesis of Schmidt~\cite{wschmidt}.

The remainder of this article is structured as follows.
In Section~\ref{sec2}, we discuss related work and how it differs from our work. 
In Section~\ref{sec3}, the threat model and the concept for this covert channel case study is proposed.
In Section~\ref{sec4}, the actual implementation of a cache-based covert channel within the context of the Genode Operating System Framework is presented.
In Section~\ref{sec5}, the covert channel implementation is evaluated and applicable countermeasures are discussed.
In Section~\ref{sec6}, we conclude the article.

\section{Related Work}\label{sec2}

The potential of covert channels to circumvent access control policies has been extensively discussed by Lampson~\cite{lampson:1973:ncp:362375.362389}, by Kemmerer~\cite{kemmerer:1983:srm:357369.357374} and by many more authors.

Cache-based covert channels, side channels and their respective countermeasures have been investigated before by Xu et al.~\cite{Xu:2011:ELC:2046660.2046670}.
While Xu et al. study covert channels based upon a CPU's L2 cache, we specifically target \emph{software-based} covert channels in the context of security-critical operating system architectures.
Thereby, we specifically address the problem of the design of secure operating system architectures.
Only by the consideration of secure design patterns for both software and hardware, the problem of covert channels within high-assurance computing systems can be effectively mitigated. 

Kong et al.~\cite{10.1109/tc.2012.78} discuss countermeasures against software-based side channel attacks.
In difference to side channels, covert channels are not based upon side effects of the computing environment, but they are actively established by attackers in order to interconnect two initially isolated processes.
Moreover, our work is not only targeted at the creation and limitation of covert channels and side channels, but it also addresses implications on the security architecture of future variants of security-critical operating systems.

The work of Wendzel~\cite{wendzeldiss} is also very relevant to the area of contemporary (network) covert channel research.
However, in contrast to the work of Wendzel, the threat model implied by this article is not based on attack patterns in the context of network policy violations.
Instead, we exploit fundamental aspects of the operating system's architecture, which should originally be designed to provide proper isolation between different compartments.
In this respect, the operating system covert channel approach presented by us can be perceived as a special variant of an \emph{insider attack}, as the attacker is assumed to have regular access to some of the operating system's processes, but he is prohibited from accessing higher privileged process by the operating system's policy.
Our considerations on this topic are especially important as, indeed, insider attacks are considered as one of the most dangerous threats to confidentiality in the context of DLP (data leakage protection)~\cite{Stamati-Koromina:2012:ITC:2371316.2371374}.

D{\"u}rmuth~\cite{duermuth09} presents important considerations on covert channels, including a novel type of covert channel within the PostScript language.
In contrast to D{\"u}rmuth, we are specifically looking at the operating system level, thus, we are targeting a slightly different branch of covert channels.

Recently, Peter et al.~\cite{cryptoeprint:2014:984} have presented a study on the possibilities of covert channels within micro kernels.
In difference to these authors, we do not really care for potential flaws in a specific micro kernel, but we are specifically targeting a covert channel that is established over central operating system components, which might be running on top of arbitrary micro kernels.

As we have now discussed the work of other authors, we will continue by presenting the threat model of this kind of covert channel attacks, and by outlining the conceptual background of the proposed covert channel.

\section{Threat Model and Concept for a Cache-based Covert Channel}\label{sec3}

The threat model of the presented covert channel can be described as follows.
An attacker wants to extract secret information from a restricted compartment within the computing system (e.g., within a multilevel security context).
The attacker is assumed to have direct access to the \emph{receiving process} as part of the covert channel interaction, but he is unable to directly access the restricted compartment (i.e., the \emph{sending process}).
By establishing the covert channel, the attacker would be enabled to indirectly access the restricted compartment, and he would be able to violate the implemented access control policy.
As the receiving process might be connected to other untrusted resource (e.g., the Internet), the attack is not restricted to local intrusions at all, but the attacker might be able to infiltrate the system from any location in the world.

Regarding the concept of this covert channel, the covert channel is established over a software-based file system cache.
Based on the file system performance, a hidden stream is modulated by the sender and evaluated by the receiver.
As this type of covert channel is essentially based on the timing behavior of the file system cache, it can also be classified as a \emph{covert timing channel}.

The sender modulates a 1-bit by reading in a file residing in its compartment and evicting the contents of the software cache.
For modulating a 0-bit, the sender performs no file system related actions and waits for a specific threshold time, which has to be known to both the sender and the receiver.
Obviously, this requires initial compatibility between sender and receiver, but this precondition applies to all covert channel technologies.

\begin{figure}[!ht]
\includegraphics[width=1.0\textwidth]{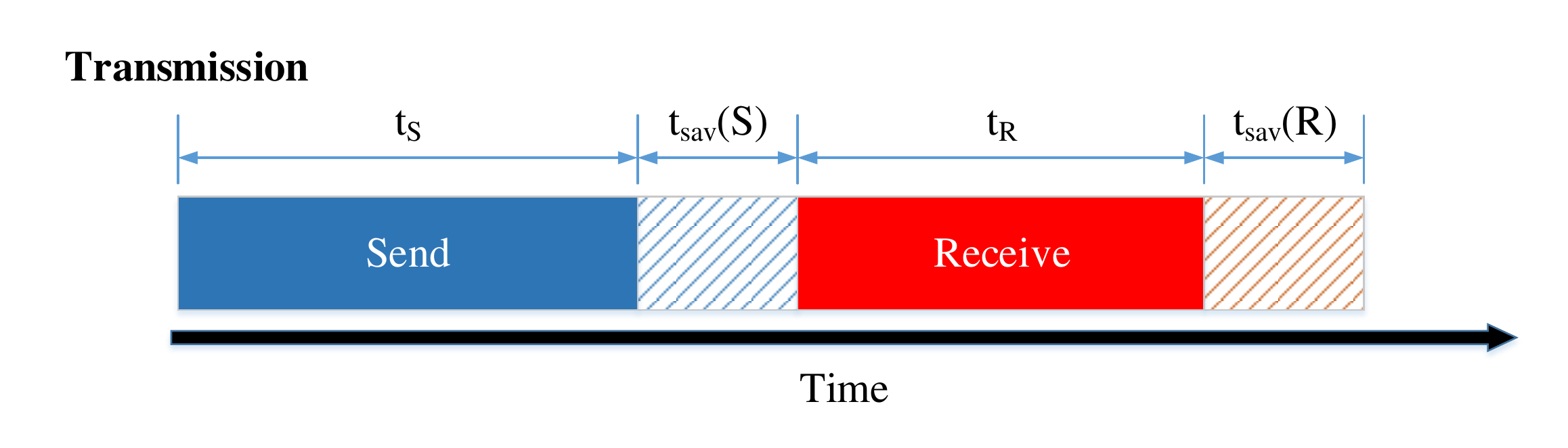}
\caption{Covert channel transmission between sender and receiver}
\label{transmission}
\end{figure}

The receiver demodulates the data by repeatedly reading in another file in its own compartment and measuring the performance of this process.
If the read-in process takes up more time than defined in a specific threshold, a 1-bit is assumed.
Following the read-in process, the receiver waits for an additional time period.
In case the amount of time needed for the read-in process is shorter than the defined threshold, a 0-bit is assumed.
Fig.~\ref{transmission} describes the covert channel transmission/reception.

The time needed for transmitting one bit is the maximum time needed by the sender for sending one bit plus the maximum time for the receiver for receiving the bit. 
Because transmissions from a permanent storage medium to RAM can slightly differ in time, both sender and receiver are in need of an additional safety time frame. 
Hence, the time $t_b$ for transmitting a bit $b$ can be estimated by calculating the individual read times of data from slow media with an additional safety time frame $t_{sav}$ for both the size of the sending frame $S$ and the size of the receiving frame $R$, as described within Eq.~\ref{eq1}.

\begin{equation}\label{eq1}
t_{b} = \left(t_{S}+t_{sav}(S)\right) + \left(t_{R}+t_{sav}(R)\right)
\end{equation}

In this context, the function $t_{sav}$ returns the safety time frame for a given frame size or transmission time.
The time needed to transmit multiple bits can be calculated by multiplying the result of Eq.~\ref{eq1} $n$ times, where $n$ is the number of bits to transmit, as shown in Eq.~\ref{eq2}.

\begin{equation}\label{eq2}
t_{total}=n * t_{b}
\end{equation}

\begin{figure}[!ht]
\includegraphics[width=1.0\textwidth]{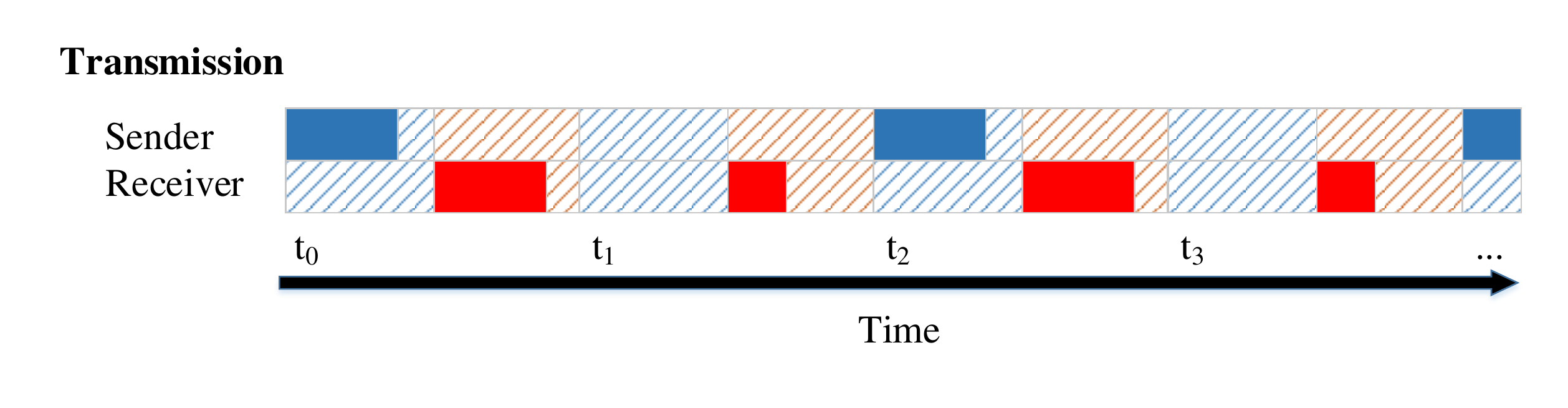}
\caption{Multiple bit transmissions with the proposed covert timing channel}
\label{multi}
\end{figure}

In other words both sender and receiver have fixed time slots for sending or receiving each bit.
Fig.~\ref{multi} shows multiple transmissions with the covert channel.
The sender sends a pattern with alternating ones and zeros (10101...).
The filled bars denote file reading periods and the striped bars represent waiting periods.
The receiver will have slightly different read times while receiving a 1-bit in comparison to receiving a 0-bit.

In the time frame starting at $t_{0}$ the sender (blue) evicts the whole file system cache. 
Consequently, the receiver (red) will measure that the time for reading his own file exceeds the previously defined threshold and recognize that a 1-bit is sent. 
In the time frame starting at $t_{1}$ the sender waits while having no file system related activity. 
Therefore, the receiver recognizes a transmitted 0-bit, as the time measured for reading his own file is found to be below the threshold.

To gain a better estimation of the length of transmission periods, the following preconditions are assumed:

\begin{enumerate}
\item The sender always evicts the whole file system cache of size $c$ (MB). 
\item The read performance of the slower media (hard drive or SSD) is $l$ (MB/s). 
\item Therefore, the sender requires $\frac{c}{l}$ seconds to evict the cache (plus the safety time frame).
\item The receiver reads two blocks from the slower media (equal to 1 MB). 
\end{enumerate}

Thus, the total time for transmitting one bit may be calculated by Eq.~\ref{eq3}.

\begin{equation}\label{eq3}
t_{b}=\left(\frac{c}{l}+t_{sav}\left(\frac{c}{l}\right)\right)+\left(\frac{1}{l}+t_{sav}\left(\frac{1}{l}\right)\right)
\end{equation}

Up to this point, the sending and receiving processes consume large portions of the transmission time for waiting periods.
However, optimization is possible, as the sender does not necessarily have to wait for the end of a whole receiving period in order to start another transmission.
Likewise, the receiver may start analyzing the transmission while it is still being sent out.
If enough data is already evicted from the cache, the receiver will already be able to measure a read time above the previously defined threshold.

As the considered cache is a block-based file system cache, it is generally sufficient to wait for the time period in which the first block of the receiver's file is evicted. 
Afterwards, the receiver just reads in its own file again in order to store the file contents within the cache. 

\begin{figure}[!ht]
\includegraphics[width=1.0\textwidth]{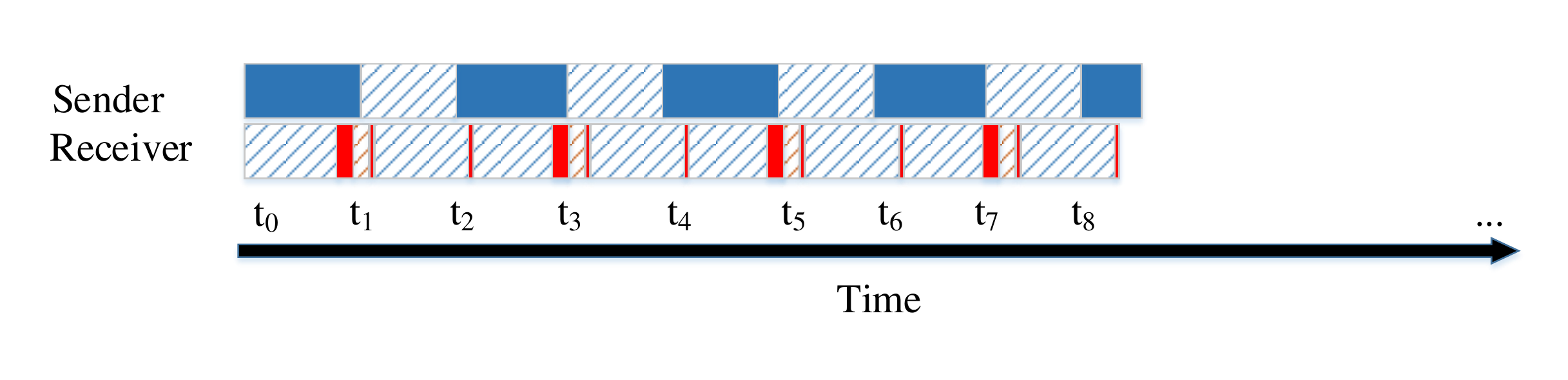}
\caption{Multiple bit transmissions with the optimized variant of the covert channel}
\label{multi_opti}
\end{figure}

The time periods of sender and receiver may jitter slightly:
If the receiver reads in its own file a little bit too late, the sender may have already evicted the cache for the transmission of the next bit.
If it reads in the file a little bit too early, the file may not yet be evicted from the cache.
The optimized version of the covert channel transmission/reception is depicted in Fig.~\ref{multi_opti}.

In Fig.~\ref{multi_opti} it is made visible that sender and receiver have a divergence.
Up to a certain point, such a divergence can be tolerated by the described algorithm, but if the divergence gets too high, the waiting period of the sender is adjusted.
The primary advantages of the optimized covert channel approach can be summarized as follows:

\begin{itemize}
\item With the optimized transmission pattern, the previously introduced safety time frames are no longer necessary, which is also formalized within Eq.~\ref{eq4}.

\begin{equation}\label{eq4}
t_{b}=\frac{c}{l}+\frac{1}{l}
\end{equation}

\item With the new transmission pattern, the receiver only needs to wait for the time period needed to receive one additional bit at the end of the transmission, i.e., the receiver needs to additionally read 1 MB (= 2 blocks) in order to receive the last bit over the covert channel.
This matter of fact is also described in Eq.~\ref{eq5}

\begin{equation}\label{eq5}
t_{total}=n * t_{b} + \frac{1}{l}
\end{equation}

\item For transmitting a 0-bit the sender does not need to wait for a whole sending period, but it can wait for a smaller, fixed-size time period.
\end{itemize}

Having discussed the theoretical background and the transmission patterns of the proposed covert timing channel, the next section will describe the specific implementation of the covert channel within the context of a security-critical operating system.

\section{Implementation of the Covert Channel}\label{sec4}

\begin{figure}[!ht]
\includegraphics[width=1.0\textwidth]{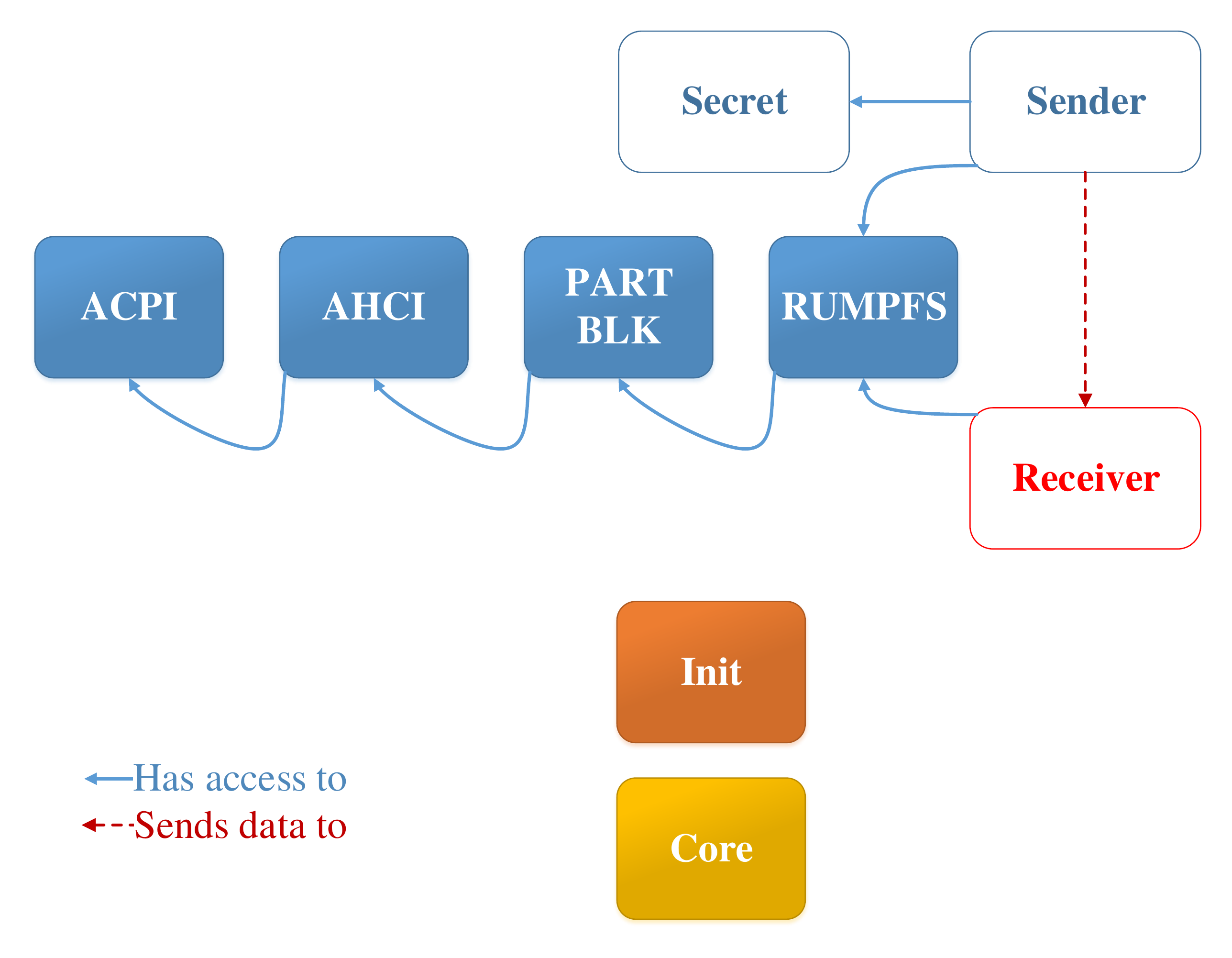}
\caption{System architecture used for this case study}
\label{arch}
\end{figure}

The proposed covert channel is based on the Genode Operating System Framework~\cite{genode}, which provides a well-established methodology to create new compositions of component-based operating systems based on different flavors of micro kernels. 
The Genode Framework will serve as an example for our experiments on covert channel establishment, but our considerations equally apply to other kinds of component-based operating systems.
In this context, the following operating system architecture (Fig.~\ref{arch}) was implemented.

Sender and receiver are distinct processes that are prohibited from any communication with each other in terms of system policy.
The sender is a higher privileged process that may access a secret, but it is not allowed to forward the secret, because it cannot write to any lower privileged processes (i.e., it is implementing the Bell-LaPadula model~\cite{bell1973}).

Both sender and receiver are connected to \emph{RUMPFS}, which is a file system service that also implements caching of blocks for all connected clients.
Both processes are explicitly prohibited from communicating with each other via RUMPFS, but they are both using the same computing resources for the caching of blocks.
While it would also be possible to use individual caches for each process, this would effectively undermine the primary performance advantages of the cache and the cache could be as well omitted.

ACPI, AHCI and PART\_BLK are system services that implement low level hardware access and partitioning of hard drives.
INIT and CORE are basic system services, which establish an interface to the underlying NOVA~\cite{Steinberg:2010:NMS:1755913.1755935} micro kernel, and which delegate access control rights and other capabilities to the overlying processes.

The implementation of the previously described covert channel effectively demonstrates how a component-based operating system with a minimized trusted computing base can still be vulnerable to attacks that are based upon very basic components, which are usually required for performance reasons, such as the utilized file system block cache.
 
After we have discussed the architecture of the proposed covert timing channel within the real world context of a component-based operating system, we will now look at the evaluation of the covert channel and at possible countermeasures that can be applied to mitigate this kind of covert channel.

\section{Evaluation and Countermeasures}\label{sec5}

We have practically demonstrated the implementation of the described covert channel between two processes within the context of the Genode Framework.
For transmissions over the covert channel, both processes are accessing distinct files, which are subject to caching within a 32 MB block cache.
A message of 27 1-bits and 37 0-bits is transmitted in order to evaluate the transmission times of the covert channel with different block cache sizes.

\begin{figure}[!ht]
\includegraphics[width=1.0\textwidth]{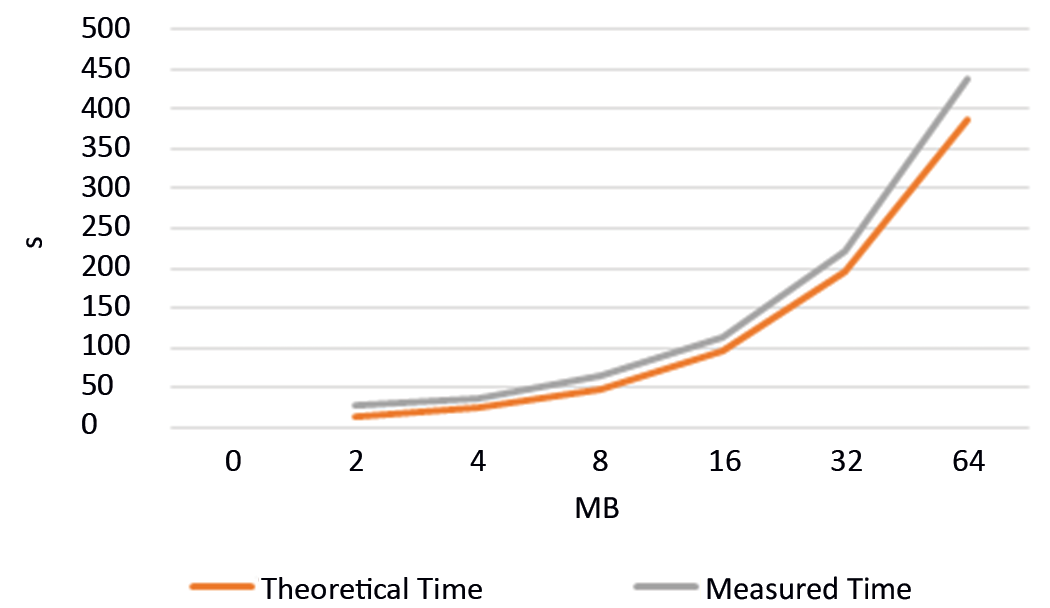}
\caption{Comparison of theoretical and actually measured transmission times with different block cache sizes}
\label{times}
\end{figure}

The theoretical transmission time for a single bit is deducted from the transmission bandwidth of the operating system's storage subsystem in consideration of the installed HDD type.
In order to compute the total transmission time, the transmission time of a single bit is multiplied with the number of transmitted 1-bits, and corresponding waiting periods are added for each 0-bit.
The actually measured times were slightly higher as computed, but the computation can be used for a general assessment of the performance of the covert channel with different block cache sizes (see also Fig.~\ref{times}).

As can be seen in Fig.~\ref{times}, the transmission time of a message is heavily depending on the size of the block cache, and the time needed for evicting the whole cache is relatively proportional to the size of the cache.
For different cache sizes, Tab.~\ref{schmidt_table} compares the theoretical times with the actually measured times, and presents the actually measured performance (bit/s).

\begin{table}[!ht]
\centering
\caption{Performance evaluation results of the proposed covert channel. Source:~\cite{wschmidt}\label{schmidt_table}}
\renewcommand{\arraystretch}{1.3}
\begin{tabular}{ | c | c | c | c | c | c | c | c | }
\hline
\vtop{\hbox{\strut Cache}\hbox{\strut Size}} & \vtop{\hbox{\strut Read}\hbox{\strut Time}} & \vtop{\hbox{\strut Wait}\hbox{\strut Period}} & \vtop{\hbox{\strut Theor.}\hbox{\strut Time}\hbox{\strut (1-Bits)}} & \vtop{\hbox{\strut Theor.}\hbox{\strut Time}\hbox{\strut (0-Bits)}} & \vtop{\hbox{\strut Theor.}\hbox{\strut Time}} & \vtop{\hbox{\strut Measured}\hbox{\strut Time}} & Performance \\
\hline
\hline
2 MB & 0.292 s & 0.125 s & 7.873 s & 4.625 s & 12.498 s & 28.238 s & 2.266 bit/s \\
\hline
4 MB & 0.566 s & 0.25 s & 15.273 s & 9.25 s & 24.523 s & 37.175 s & 1.721 bit/s \\
\hline
8 MB & 1.122 s & 0.5 s & 30.014 s & 18.5 s & 48.514 s & 65.500 s & 0.977 bit/s \\
\hline
16 MB & 2.224 s & 1.0 s & 60.055 s & 37.0 s & 97.055 s & 113.543 s & 0.564 bit/s \\
\hline
32 MB & 4.473 s & 2.0 s & 120.762 s & 74.0 s & 194.762 s & 221.501 s & 0.289 bit/s \\
\hline
64 MB & 8.810 s & 4.0 s & 237.866 s & 148.0 s & 385.866 s & 436.724 s & 0.147 bit/s \\
\hline
\end{tabular}
\end{table}

Because the theoretical times are only estimated values based upon general storage performance values, theoretical and measured times differ, and the measured performance was marginally lower.
With a cache size of 2 MB, the proposed covert channel is able to reliably deliver a performance of ~2 bit/s.
Although this number seems small, the covert channel is absolutely fast enough to deliver small-sized messages such as passwords and cryptographic keys between sender and receiver.

\begin{figure}[!ht]
\includegraphics[width=1.0\textwidth]{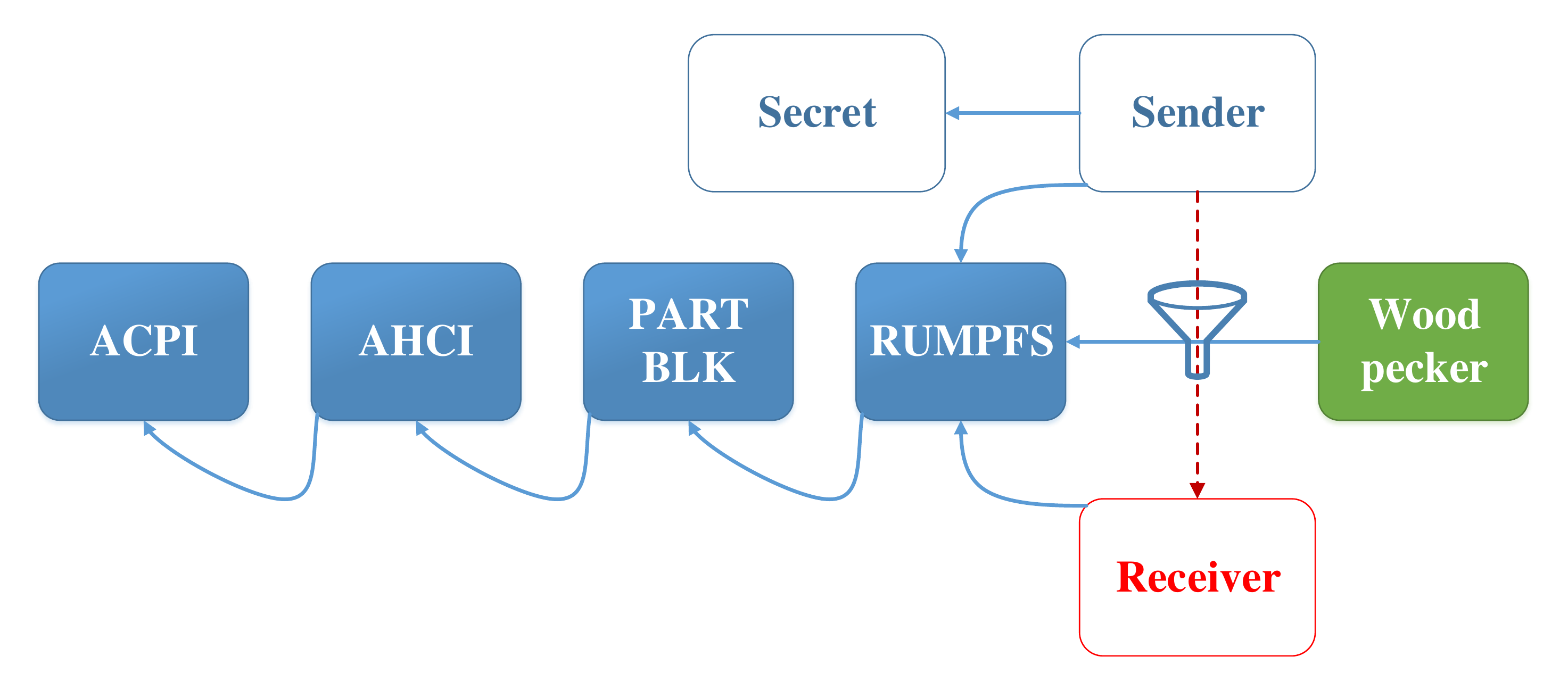}
\caption{Mitigating the covert channel with the help of a dedicated disrupting process}
\label{woodpecker}
\end{figure}

Regarding countermeasures against the implemented covert channel, we have been able to disrupt the implemented covert channel by clearing the cache in fixed intervals.
The process of disrupting the covert channel transmission (called \emph{woodpecker}) is depicted in Fig.~\ref{woodpecker}.

The disrupting process works analogous to the sending process.
A specific file is read in repeatedly and, thereby, previously stored contents are evicted from the cache.
By this behavior, a normal process accessing the cache is only minimally influenced, but the covert channel suffers in two ways: 
The receiver will notice additional 1-bits as the read-in period is longer whenever the disrupting process is also active, and the timing between sender and receiver is deferred, as the sender also needs more time for reading in its own file. 

A successful method to harden the channel against disruptions is to use a (3,1)-Hamming-Code, which is using three bits instead of one to send a single bit of the original message.
The result of a bit transmission as part of the original message is determined by analyzing the bit pattern of three consecutively transmitted bits, i.e., a pattern of 000, 001, 010, 100 is recognized as a 0-bit, and 011, 101, 110, 111 is recognized as a 1-bit.
Consequently, each transmission with the hardened covert channel demands three times of the size and transmission time of the original message.

While it would still be possible to disrupt the hardened covert channel by automatically disrupting the transmission of every single bit, this would also diminish the performance benefits of using a file system block cache in the first place.
Furthermore, this methodology might also add a constant interval to the file reading performance, and the covert channel could still be established by taking this fixed interval into account for each bit transmission.

Thus, only if shared caches would be completely removed from the operating system architecture, the possibility of such a type of covert channel could be completely precluded at the cost of a reduced storage access performance.
Moreover, the underlying hardware would still enable the establishment of potential covert channels, although specific hardware cache designs (see Wang and Lee~\cite{Wang:2007:NCD:1250662.1250723}) are able to provide some degree of non-interference between processes accessing the same cache.
Finally, it would be possible to mitigate covert channel attacks via specific \emph{guard} processes (see Alves-Foss et al.~\cite{alves-foss06themils}) that have to be configured to detect irregularities in cache access patterns.

The next section will conclude the article and provide an outlook on possible opportunities for future research.

\section{Conclusion}\label{sec6}

The implementation of the proposed covert timing channel reveals important criteria for the implementation of security-driven operating systems architectures. 
In difference to many other kinds of covert channel attacks, our approach is solely based upon software means.
Because of this, the covert channel is not depending on any specific type of CPU architecture, but it can be applied to any component-based operating system with a similar storage processing architecture as described within the article.
Using a file system-based software cache, we are able to transfer a short message over the covert channel. 
While the covert channel can be actively disrupted by other processes, we have been able to stabilize the covert channel by utilizing a (3,1)-Hamming code.
Exhaustive information on the performance and feasibility of the covert channel can be found in the master's thesis of Schmidt~\cite{wschmidt}.

In summary, we are able to demonstrate that the implemented covert channel is capable of reliably transferring pieces of information from one compartment to another compartment.
Thereby, we are able to effectively circumvent the operating system's security policy, and to perform a specific type of insider attack.

Future research work might include further improvements on the efficiency of the covert channel, new means for disrupting the channel, and more applications of similar covert channels.

\bibliographystyle{abbrv}
\bibliography{paper_covchan}  

\begin{thebibliography}{10}

\bibitem{alves-foss06themils}
J.~Alves-Foss, S.~W. Harrison, P.~Oman, and C.~Taylor.
\newblock {The MILS Architecture for High-Assurance Embedded Systems}.
\newblock {\em International Journal of Embedded Systems}, 2(4):239--247, Dec.
  2006.

\bibitem{bell1973}
D.~E. Bell and L.~J. LaPadula.
\newblock {Secure Computer Systems: Mathematical Foundations}.
\newblock {\em The MITRE Corporation, Tech. Rep. 2547}, I, Mar. 1973.

\bibitem{duermuth09}
M.~D{\"u}rmuth.
\newblock {\em {Novel classes of side channels and covert channels}}.
\newblock PhD thesis, Saarland University, Saarbr{\"u}cken, Germany, Dec. 2009.

\bibitem{Feamster:2002:ICW:647253.720281}
N.~Feamster, M.~Balazinska, G.~Harfst, H.~Balakrishnan, and D.~Karger.
\newblock {Infranet: Circumventing Web Censorship and Surveillance}.
\newblock In {\em Proceedings of the 11th USENIX Security Symposium}, pages
  247--262, Berkeley, CA, USA, Aug. 2002. USENIX Association.

\bibitem{genode}
{Genode Labs}.
\newblock {Genode Operating System Framework}.
\newblock Project Documentation, \url{http://www.genode.org}.

\bibitem{Heimdahl98formalmethods}
M.~P.~E. Heimdahl and C.~L. Heitmeyer.
\newblock {Formal Methods For Developing High Assurance Computer Systems:
  Working Group Report}.
\newblock In {\em Proceedings of the Second IEEE Workshop on
  Industrial-Strength Formal Techniques}, pages 1--5. IEEE, Oct. 1998.

\bibitem{jaeger:1998:sac:319195.319229}
T.~Jaeger, J.~Liedtke, V.~Panteleenko, Y.~Park, and N.~Islam.
\newblock {Security Architecture for Component-Based Operating Systems}.
\newblock In {\em Proceedings of the 8th ACM SIGOPS European Workshop on
  Support for Composing Distributed Applications}, EW 8, pages 222--228, New
  York, NY, USA, Sept. 1998. ACM.

\bibitem{kemmerer:1983:srm:357369.357374}
R.~A. Kemmerer.
\newblock {Shared Resource Matrix Methodology: An Approach to Identifying
  Storage and Timing Channels}.
\newblock {\em ACM Trans. Comput. Syst.}, 1(3):256--277, Aug. 1983.

\bibitem{10.1109/tc.2012.78}
J.~Kong, O.~Aciicmez, J.-P. Seifert, and H.~Zhou.
\newblock {Architecting against Software Cache-Based Side-Channel Attacks}.
\newblock {\em IEEE Transactions on Computers}, 62(7):1276--1288, July 2013.

\bibitem{lampson:1973:ncp:362375.362389}
B.~W. Lampson.
\newblock {A Note on the Confinement Problem}.
\newblock {\em Commun. ACM}, 16(10):613--615, Oct. 1973.

\bibitem{cryptoeprint:2014:984}
M.~Peter, J.~Nordholz, M.~Petschick, J.~Danisevskis, J.~Vetter, and J.-P.
  Seifert.
\newblock {Undermining Isolation through Covert Channels in the Fiasco.OC
  Microkernel}.
\newblock Cryptology ePrint Archive, Report 2014/984, Dec. 2014.
\newblock \url{http://eprint.iacr.org/2014/984}.

\bibitem{wschmidt}
W.~Schmidt.
\newblock {Covert Channels und Schutzmaßnahmen beim Festplatten-Blockcache im
  Genode Operating System Framework (German language content)}.
\newblock {Master's Thesis}, FernUniversit{\"a}t in Hagen, Germany, Oct. 2014.

\bibitem{Stamati-Koromina:2012:ITC:2371316.2371374}
V.~Stamati-Koromina, C.~Ilioudis, R.~Overill, C.~K. Georgiadis, and
  D.~Stamatis.
\newblock {Insider Threats in Corporate Environments: A Case Study for Data
  Leakage Prevention}.
\newblock In {\em Proceedings of the Fifth Balkan Conference in Informatics},
  BCI '12, pages 271--274, New York, NY, USA, Sept. 2012. ACM.

\bibitem{Steinberg:2010:NMS:1755913.1755935}
U.~Steinberg and B.~Kauer.
\newblock {NOVA: A Microhypervisor-based Secure Virtualization Architecture}.
\newblock In {\em Proceedings of the 5th European Conference on Computer
  Systems}, EuroSys '10, pages 209--222, New York, NY, USA, Apr. 2010. ACM.

\bibitem{Wang:2007:NCD:1250662.1250723}
Z.~Wang and R.~B. Lee.
\newblock {New Cache Designs for Thwarting Software Cache-based Side Channel
  Attacks}.
\newblock In {\em Proceedings of the 34th Annual International Symposium on
  Computer Architecture}, ISCA '07, pages 494--505, New York, NY, USA, 2007.
  ACM.

\bibitem{wendzeldiss}
S.~Wendzel.
\newblock {\em {Novel Approaches for Network Covert Storage Channels}}.
\newblock PhD thesis, FernUniversit{\"a}t in Hagen, Faculty for Mathematics and
  Computer Science, Hagen, Germany, May 2013.

\bibitem{Xu:2011:ELC:2046660.2046670}
Y.~Xu, M.~Bailey, F.~Jahanian, K.~Joshi, M.~Hiltunen, and R.~Schlichting.
\newblock {An Exploration of L2 Cache Covert Channels in Virtualized
  Environments}.
\newblock In {\em Proceedings of the 3rd ACM Workshop on Cloud Computing
  Security Workshop}, CCSW '11, pages 29--40, New York, NY, USA, 2011. ACM.

\end{thebibliography}

\end{document}